\documentclass[prl,twocolumn,showpacs,amsmath,amssymb,citeautoscript]{revtex4}

\usepackage{graphicx,times}

\begin{document}

\title{Partition Functions of Strongly Correlated Electron Systems as ``Fermionants'' }
\author{Shailesh Chandrasekharan$^1$ and Uwe-Jens Wiese$^2$}
\affiliation{$^1$Department of Physics, Box 90305, Duke University
Durham, North Carolina 27708, USA. \\
$^2$Albert Einstein Center for Fundamental Physics,
Institute for Theoretical Physics, Bern University,
Sidlerstrasse 5, CH-3012, Bern, Switzerland}
\begin{abstract}
We introduce a new  mathematical object, the ``fermionant'' ${\mathrm{Ferm}}_N(G)$, of type $N$ of an $n \times n$ matrix $G$. It represents certain $n$-point functions involving $N$ species of free fermions. When $N=1$, the fermionant reduces to the determinant. The partition function of the repulsive Hubbard model, of geometrically frustrated quantum antiferromagnets, and of Kondo lattice models can be expressed as fermionants of type $N=2$, which naturally incorporates infinite on-site repulsion. A computation of the fermionant in polynomial time would solve many interesting fermion sign problems.
\end{abstract}


\maketitle


The Pauli exclusion principle has profound physical consequences, ranging from the stability of condensed matter to the structure of atoms, nuclei, as well as of protons and neutrons. In quantum mechanics, the Pauli principle manifests itself through the requirement that the wave function must be antisymmetric under the exchange of two fermions. In many-body quantum mechanics this is incorporated by the fact that fermion creation and annihilation operators anti-commute with one another. When a fermionic path integral is set up in an occupation number basis, the resulting worldlines of identical fermions form closed loops wrapping around the periodic Euclidean time direction. In relativistic theories, these worldlines can also form closed loops which incorporate the physics of pair creation and annihilation. The number of closed loops, which corresponds to the number of cycles in a permutation of fermion positions, determines the fermion permutation sign of a configuration of worldlines. Alternatively, fermion path integrals can be constructed in a basis of coherent states using anti-commuting Grassmann variables. Grassmann variables can be viewed as a different way of bookkeeping of the fermion worldlines. Indeed, Feynman diagrams are well-known representations of these worldlines and the negative sign associated with each fermion loop is nothing but the fermion permutation sign.

While worldlines of bosonic particles can be given a classical interpretation, the same is not true for fermions due to the fermion permutation sign, which is an important quantum mechanical phase that often makes the weight of a configuration of fermion worldlines negative. This makes it impossible to interpret this weight as a probability in a Monte Carlo simulation based on importance sampling. Fortunately, it is possible to sum over all fermion worldlines when fermions are not interacting with one another. For this reason, the traditional aim of Monte Carlo approaches to fermions is to convert an interacting into a non-interacting problem, by introducing auxiliary fields which incorporate the interactions. Once the fermions are freely moving in the auxiliary field, one can sum over all fermion worldlines which leaves behind a fermion determinant as the weight of each auxiliary field configuration.  In cases where this determinant is positive, the sign problem is considered solved and a Monte Carlo algorithm can be devised to sample the auxiliary field. This has been the method of choice for almost all strongly correlated fermion problems until recently. However, in general there is no guarantee that the fermion determinant is positive. The physics of the auxiliary field plays an important role in determining its sign. There are many interesting quantum systems, ranging from high-temperature superconductors, geometrically frustrated antiferromagnets, and Kondo lattices, to systems of quarks and gluons at non-zero baryon density, where indeed the fermion determinant can be negative or complex. We refer to these as difficult sign problems that remain unsolved. If only the magnitude of the weight of a configuration is used for importance sampling while its sign or complex phase is included in measured observables, the error to signal ratio and thus the required computational effort increases exponentially with the space-time volume. 

Over the last decade new ways to sum over fermion worldline configurations have emerged which allow us to write the partition function as a sum over novel types of configurations with non-negative weights. In the meron-cluster approach, fermion worldlines are divided into independent clusters. A summation over the fermionic degrees of freedom within a cluster produces a non-negative weight for each cluster configuration \cite{PhysRevLett.83.3116,Chandrasekharan:2002vk}. More recently, this idea was extended to the notion of fermion bags, where again a summation of fermion worldlines within a bag yields a non-negative weight for each fermion bag \cite{Chandrasekharan:2009wc}. In yet another development, in certain models a sum over all Feynman diagrams of a given order in the perturbative expansion can be performed and yields a non-negative weight for these diagrams. This leads to the diagrammatic determinantal Monte Carlo method \cite{PhysRevB.72.035122, RevModPhys.83.349}. While these new approaches allow us to solve some sign problems, which seemed unsolvable with the traditional auxiliary field method, none of these new developments have yet solved some of the difficult sign problems mentioned above.

In this letter we further investigate the nature of these difficult fermion sign problems by showing that the partition function of some strongly correlated electron systems, including the infinite $U$ Hubbard model away from half filling, the quantum Heisenberg model for geometrically frustrated antiferromagnets, as well as the Kondo lattice system, can be expressed in terms of a new mathematical object --- which we name the ``fermionant'' ---  which is a generalization of the determinant of a matrix. The exact evaluation of the partition function is a sum over exponentially many states in the Hilbert space and so is generically impossible in polynomial time. Remarkably, for free fermions, since the partition function is a determinant  of an $n \times n$ matrix it can be computed with an effort proportional to $n^3$. Our results imply that some interesting strongly correlated electron systems share certain features with free fermions and that in these systems the determinant is replaced by a fermionant. However, it is not clear whether the fermionant can be computed in polynomial time. Thus, we think that perhaps the computation of a fermionant is at the heart of the difficult sign problems. 
Fermionants naturally arise in systems in which different fermion species (such as spin up and spin down) are not allowed to simultaneously occupy the same lattice point. Interestingly, fermionants also describes a class of  $n$-point functions in theories of free fermions. 


Let $G$ be an arbitrary real- or complex-valued $n \times n$ matrix. The fermionant of type $N$ of the matrix $G$ is defined as
\begin{equation}
{\mathrm{Ferm}_N(G)} = (-1)^n \sum_{\sigma} (- N)^{\nu(\sigma)} G_{1\sigma(1)} G_{2\sigma(2)} \dots G_{n\sigma(n)},
\end{equation}
where the sum extends over all permutations $\sigma$ of the indices 1,2,\dots,$n$, $\nu(\sigma)$ is the number of cycles in the permutation, and $\sigma(i)$ is the permutation partner of the index $i$. Based on this definition, it is clear that the fermionant of type $N=1$ is simply the determinant. 

Let us consider $N$ species of fermions moving on a space-time lattice with sites labeled by the index $x=1,2,\dots,V$, where $V$ is the space-time volume. We use $\psi_{x,\alpha}$ and $\overline\psi_{x,\alpha}$ to denote the Grassmann variables associated with the species $\alpha=1,2,\dots,N$ of the fermion at the site $x$. We assume that the fermions hop on the space-time lattice as non-interacting particles, based on the $U(N)$-invariant Euclidean action
\begin{equation}
\label{action}
S[\overline\psi,\psi] = \sum_{\alpha = 1}^N \sum_{x,y} \overline\psi_{x,\alpha} M_{xy}
\psi_{y,\alpha},
\end{equation}
which results in the partition function
\begin{equation}
Z = \int \prod_{x,\alpha} [d\overline\psi_{x,\alpha}\ d\psi_{x,\alpha}] \exp(- S[\overline\psi,\psi]).
\end{equation}
We now define the $n$-point function
\begin{eqnarray}
C(x_1,x_2,\dots,x_n) &=& \frac{1}{Z} \int \ \prod_{x,\alpha} [d\overline\psi_{x,\alpha}\ d\psi_{x,\alpha}]
\nonumber \\
\times \ \overline\psi_{x_1}\psi_{x_1}\ &\dots& \ \overline\psi_{x_n}\psi_{x_n}\ 
\exp(- S[\overline\psi,\psi]).
\end{eqnarray}
In this expression $\overline\psi_{x_i}\psi_{x_i} = \sum_\alpha \overline\psi_{x_i,\alpha}\psi_{x_i,\alpha}$ contains a summation over the species index which has been suppressed for convenience. Using Feynman rules, one can show that $C(x_1,x_2,\dots,x_n) = \mathrm{Ferm}_N(G)$, where $G$ is the $n \times n$ matrix composed of elements connecting sites $x_i$ and $x_j$ given by $G_{x_i x_j} = (M^{-1})_{x_i x_j}$, where $M^{-1}$ is the free fermion propagator.


It is well-known that free fermion partition functions can be written as determinants of $V \times V$ matrices. It is also known that partition functions of interacting theories are sums over different $n$-point functions of a free theory. This can easily be seen in perturbation theory by expanding in powers of the coupling and then evaluating each term as a correlation function in a free theory. For some familiar strongly correlated electron models, we now show that the partition function can be expressed as a single $V$-point function of a free theory, which turns out to be a fermionant of type $N=2$ of a $V \times V$ matrix.

Before discussing the partition functions, it may be useful to recall the coherent state path integral representation of fermions using Grassmann variables. Let $c^\dagger_{\vec{x},\alpha}$ and $c_{\vec{x},\alpha}$ represent the creation and annihilation operators for electrons of spin $\alpha=\uparrow,\downarrow$ at the spatial lattice site $\vec{x}$. Introducing Grassmann variables $\psi_{\vec{x},\alpha}$ and $\overline{\psi}_{\vec{x},\alpha}$, an electron coherent state at the site $\vec{x}$ is defined as
\begin{eqnarray}
|\psi_{\vec{x}}\rangle &=& 
\exp(\psi_{\vec{x}\uparrow} c^\dagger_{\vec{x}\uparrow} + \psi_{\vec{x}\downarrow} c^\dagger_{\vec{x}\downarrow}) |0\rangle,
\nonumber \\
\langle \overline{\psi}_{\vec{x}} | &=& \langle 0 |\exp(c_{\vec{x}\uparrow}\overline\psi_{\vec{x}\uparrow} +
c_{\vec{x}\downarrow}\overline\psi_{\vec{x}\downarrow}).
\end{eqnarray}
It is easy to verify that $\langle \overline{\psi}_{\vec{x}}|\psi_{\vec{x}}\rangle = \exp(\overline{\psi}_{\vec{x},\uparrow}\psi_{\vec{x},\uparrow} + \overline{\psi}_{\vec{x},\downarrow}\psi_{\vec{x},\downarrow})$ and that the completeness relation is given by
\begin{eqnarray}
|0\rangle\langle 0| + 
\! \! c^\dagger_{\vec{x}\uparrow}|0\rangle\langle 0|c_{\vec{x}\uparrow} +
c^\dagger_{\vec{x}\downarrow}|0\rangle\langle 0|c_{\vec{x}\downarrow} +
c^\dagger_{\vec{x}\uparrow}c^\dagger_{\vec{x}\downarrow}|0\rangle\langle 0|c_{\vec{x}\downarrow}c_{\vec{x}\uparrow}
\nonumber \\
= \int \prod_\alpha [d \overline{\psi}_{\vec{x},\alpha} d \psi_{\vec{x},\alpha}]\ 
\exp(-\overline{\psi}_{\vec{x}}\psi_{\vec{x}})\ \ 
|\psi_{\vec{x}}\rangle \langle \overline{\psi}_{\vec{x}} |,
\end{eqnarray}
where we have defined $\overline\psi_{\vec{x}}\psi_{\vec{x}} = \overline{\psi}_{\vec{x},\uparrow}\psi_{\vec{x},\uparrow} + \overline{\psi}_{\vec{x},\downarrow}\psi_{\vec{x},\downarrow}$. If one wants to remove the doubly occupied state from the Hilbert space, which implies an infinitely strong repulsion between the electrons of spin up and spin down, this is easily accomplished using
\begin{eqnarray}
\hspace{-2cm} |0\rangle\langle 0| +  
c^\dagger_{\vec{x}\uparrow}|0\rangle\langle 0|c_{\vec{x}\uparrow} +
c^\dagger_{\vec{x}\downarrow}|0\rangle\langle 0|c_{\vec{x}\downarrow} = 
\nonumber \\
\int  \prod_\alpha [d \overline{\psi}_{\vec{x},\alpha} d \psi_{\vec{x},\alpha}]
(- \overline{\psi}_{\vec{x}}\psi_{\vec{x}})  
\exp(- \frac{1}{2} \overline{\psi}_{\vec{x}}\psi_{\vec{x}}) 
|\psi_{\vec{x}}\rangle \langle \overline{\psi}_{\vec{x}} |.
\label{eq:CR1}
\end{eqnarray}
If one also wants to forbid the completely empty state, one can accomplish this using the relation

\begin{eqnarray}
c^\dagger_{\vec{x}\uparrow}|0\rangle\langle 0|c_{\vec{x}\uparrow} \ +\ 
c^\dagger_{\vec{x}\downarrow}|0\rangle\langle 0|c_{\vec{x}\downarrow} = &&
\nonumber \\
\int \prod_\alpha [d \overline{\psi}_{\vec{x},\alpha} d \psi_{\vec{x},\alpha}]\ 
& (- \overline{\psi}_{\vec{x}}\psi_{\vec{x}})&
|\psi_{\vec{x}}\rangle \langle \overline{\psi}_{\vec{x}} |.
\label{eq:CR2}
\end{eqnarray}
Thus, by introducing different types of completeness relations, one can reduce the Hilbert space of the non-interacting theory and thereby introduce interactions. The partition functions of the resulting constrained theories are then equivalent to a single correlation function of a non-interacting theory. Based on this insight we now consider specific partition functions.

Let us consider the lattice action of free non-relativistic electrons hopping on a space-time lattice given by
\begin{eqnarray}
S_{\mathrm{Hubbard}} = - \sum_{\vec{x},\tau,\alpha} \Bigg\{\ \mathrm{e}^{\mu \varepsilon}
\overline\psi_{\vec{x},\tau+1,\alpha}\psi_{\vec{x},\tau,\alpha} - \frac{1}{2} \overline\psi_{\vec{x},\tau,\alpha}\psi_{\vec{x},\tau,\alpha}
\nonumber \\
\! \! \! \! \!  + \ \varepsilon t \ \mathrm{e}^{\mu\varepsilon}\ \sum_{\vec{i}} 
\Big(\overline\psi_{\vec{x},\tau+1,\alpha}\psi_{\vec{x}+\vec{i},\tau,\alpha} +\  
\overline\psi_{\vec{x}+\vec{i},\tau+1,\alpha}\psi_{\vec{x},\tau,\alpha}\Big) \Bigg\}
\end{eqnarray}
where $(\vec{x},\tau)$ is a Euclidean space-time lattice site and $\psi_{\vec{x},\tau,\alpha}$ and $\overline{\psi}_{\vec{x},\tau,\alpha}$ each represent Grassmann variables associated with the electron of spin $\alpha$ at the corresponding space-time point. The parameter $\varepsilon$ represents the lattice spacing in Euclidean time, while $t$ and $\mu$ are the hopping parameter and the chemical potential, respectively. To obtain the Hamiltonian version of the problem, one needs to take the $\varepsilon \rightarrow 0$ limit, keeping $\varepsilon L_t = 1/T$ fixed, where $T$ is the temperature and $L_t$ is the number of lattice points in Euclidean time. Based on the previous discussion of the completeness relations of coherent states, it is easy to show that the $V$-point function
\begin{equation}
Z = \int [d\overline\psi\ d\psi] \ 
\prod_{\vec{x},\tau} \Bigg(-\overline\psi_{\vec{x},\tau}\psi_{\vec{x},\tau}\Bigg) \exp(-S_{\mathrm{Hubbard}})
\label{eq:PF}
\end{equation}
is the partition function of the Hubbard model with infinite repulsion $U$ between spin up and spin down electrons. Note that we have used eq.~(\ref{eq:CR1}) just to project out the doubly occupied state from a free electron theory.

At half-filling, the infinite $U$ Hubbard model naturally reduces to a spin $\tfrac{1}{2}$ quantum antiferromagnet, provided contributions proportional to $t^2$ are maintained. In the coherent state path integral this is accomplished if one uses the action
\begin{eqnarray}
\hspace{-1cm}
S_{\mathrm{magnet}} = - \sum_{\vec{x},\tau,\alpha} \Bigg\{\ 
\overline\psi_{\vec{x},\tau+1,\alpha}\psi_{\vec{x},\tau,\alpha}
+\ 
\nonumber \\
\sqrt{\epsilon J}
\sum_{\vec{i}}
\Big(\overline\psi_{\vec{x},\tau+1,\alpha}\psi_{\vec{x}+\vec{i},\tau,\alpha} 
+ \overline\psi_{\vec{x}+\vec{i},\tau+1,\alpha}\psi_{\vec{x},\tau,\alpha}\Big) \Bigg\},
\end{eqnarray}
instead of $S_{\mathrm{Hubbard}}$ in eq.~(\ref{eq:PF}). Note that the spatial lattice remains unspecified and could hence be either bi-partite or geometrically frustrated.

Finally, a model relevant for the physics of heavy fermions is the so-called Kondo lattice model \cite{RevModPhys.69.809}. In this model a bath of free electrons interacts with a lattice of localized spins or magnetic moments. We construct this model by adding two free fermion actions. The free electron action is given by
\begin{eqnarray}
S_{\mathrm{electron}} = - \sum_{\vec{x},\tau,\alpha} \Bigg\{\ \mathrm{e}^{\mu \varepsilon}
\overline\psi_{\vec{x},\tau+1,\alpha}\psi_{\vec{x},\tau,\alpha} - \overline\psi_{\vec{x},\tau,\alpha}\psi_{\vec{x},\tau,\alpha}
\nonumber \\
+ \varepsilon t \ \mathrm{e}^{\mu\varepsilon}\ \sum_{\vec{i}} 
\Big(\overline\psi_{\vec{x},\tau+1,\alpha}\psi_{\vec{x}+\vec{i},\tau,\alpha} +  
\overline\psi_{\vec{x}+\vec{i},\tau+1,\alpha}\psi_{\vec{x},\tau,\alpha}\Big)\Bigg\},
\end{eqnarray}
where double occupancy is now allowed. Additional localized spins are described by other Grassmann fields $\chi_{\vec{R},\tau,\alpha}$ and $\overline\chi_{\vec{R},\tau,\alpha}$ associated with the lattice sites $\vec{R}$ on which impurities are located. We denote the space-time volume occupied by the impurity sites by $V_R$. The free impurity action is given by
\begin{eqnarray}
S_{\mathrm{impurity}} = - \sum_{\vec{R},\tau,\alpha} \Bigg\{\ 
\overline\chi_{\vec{R},\tau+1,\alpha}\chi_{\vec{R},\tau,\alpha} +
\nonumber \\
\sqrt{\varepsilon J} \  
\Big(\overline\chi_{\vec{R},\tau+1,\alpha}\psi_{\vec{R},\tau,\alpha} \ +\   
\overline\psi_{\vec{R},\tau+1,\alpha}\chi_{\vec{R},\tau,\alpha}\Big) \Bigg\}.
\end{eqnarray}
The partition function of the Kondo lattice model can now be written as the $V_R$-point function
\begin{eqnarray}
Z&=&\int [d\overline\psi\ d\psi] \ [d\overline\chi d\chi]\  
\prod_{\vec{R},\tau} \Bigg(-\overline\chi_{\vec{R},\tau}\chi_{\vec{R},\tau}\Bigg)\ 
\nonumber \\
&& \ \ \ \exp(-S_{\mathrm{electron}} - S_{\mathrm{impurity}})
\end{eqnarray}
Since the localized impurities can either be spin up or spin down, we now eliminate the doubly empty and doubly occupied states only of the impurity spins.

In the three cases discussed above, the partition function is a $V$-point (or in the case of the Kondo lattice model a $V_R$-point) function of a free  theory, which is given by a fermionant of type $N=2$,
\begin{equation}
Z = Z_0 \, {\mathrm{Ferm}_2(G)}, 
\end{equation}
of a $V \times V$ (or a $V_R \times V_R$) matrix $G = M^{-1}$, which represents the fermion propagator of the corresponding free fermion lattice action written in the form of eq.~(\ref{action}). The factor $Z_0 = {\mathrm{Det}(M)}^2$ is the partition function of the free lattice action. Some observables, such as certain correlation functions of fermion bilinears, can also be expressed as fermionants.


By rewriting a well-known NP-complete problem as a spin model with a sign problem, it has been shown that some sign problems are as hard as NP-complete problems \cite{Troyer:2004ge}. These problems belong to the hardest problems in the complexity class NP \cite{Mertens00}. Since it is generally expected that NP $\neq$ P, this result implies that a generally applicable Monte Carlo method that solves all sign problems with autocorrelation times that scale at most polynomially with the system size should not exist. This does not exclude that specific sign problems can still be solved. It is important to understand whether the physically relevant difficult sign problems mentioned before are also NP-complete, or whether one may hope for a solution in polynomial time. If we consider a sign problem solved when the partition function can be expressed as a sum over configurations with non-negative weights, such that the weight of each configuration can be computed in a time that grows at most polynomially with the system size, then our work shows that the solution of difficult sign problems is closely related to the computational complexity of fermionants.

Littlewood and Richardson have introduced the concept of the immanant 
\begin{equation}
{\mathrm{Imm}}_\Gamma(G) = \sum_{\sigma} \chi_\Gamma(\sigma) 
G_{1\sigma(1)} G_{2\sigma(2)} \dots G_{n\sigma(n)},
\end{equation}
as a generalization of the determinant of an $n \times n$ matrix $G$ \cite{Littlewood34}. Here $\chi_\Gamma(\sigma)$ is the character of the permutation group element $\sigma$ in the representation $\Gamma$ (characterized by a Young tableau with $n$ boxes). When one considers the totally antisymmetric representation (associated with the Young tableau consisting of a single column of $n$ boxes), the immanant reduces to the determinant. Similarly, when one considers the totally symmetric representation (associated with the Young tableau consisting of a single row of $n$ boxes), the immanant reduces to the permanent ${\mathrm{Perm}}_\Gamma(G) = \sum_{\sigma}  G_{1\sigma(1)} G_{2\sigma(2)} \dots G_{n\sigma(n)}$. The evaluation of the permanent is even more difficult than solving NP-complete problems \cite{Valiant79}, and the computational complexity of a general immanent is similar to the one of the permanent \cite{Barvinok90}. Remarkably, thanks to its invariance properties, the closely related determinant, whose naive computation  would require an effort proportional to $n!$, can be calculated in just $n^3$ steps. 

Just like the permanent and other immanants, the fermionant of a matrix is a generalization of the determinant. The fermionant weighs the contribution of a permutation $\sigma$ with the factor $(- N)^{\nu(\sigma)}$, where $\nu(\sigma)$ is the number of cycles of $\sigma$. Just like the character $\chi_\Gamma(\sigma)$, the number of cycles
$\nu(\sigma)$ is the same for all members of a conjugacy class. The characters form a complete orthogonal system. Consequently, one can expand
\begin{equation}
(- N)^{\nu(\sigma)} = \sum_\Gamma c_\Gamma(N) \chi_\Gamma(\sigma),
\end{equation}
such that the fermionant can be expressed as a linear combination of immanants
\begin{equation}
{\mathrm{Ferm}}_N(G) = (-1)^n \sum_\Gamma c_\Gamma(N) {\mathrm{Imm}}_\Gamma(G).
\end{equation}
While this relation is not useful for numerical purposes, it shows that the fermionant is mathematically intimately related to the immanants. Although no polynomial time algorithm is known for the evaluation of a general immanant, this does not necessarily mean that the search for a polynomial time algorithm to evaluate the fermionant is doomed from the outset. In particular, a careful examination of its invariance properties seems worthwhile. Even if the evaluation of a general fermionant may require a non-polynomial computational effort, some fermionants can still be sampled stochastically in polynomial time. For example, non-frustrated quantum antiferromagnets, whose partition functions indeed are fermionants, can be simulated with the very efficient loop-cluster algorithm \cite{Evertz:1992rb,Wiese:1994ab}. Since it arises in several important physical systems, a careful study of the mathematical properties of the fermionant is highly desirable.


In conclusion, we have investigated systems of strongly interacting fermions including the infinite $U$ Hubbard model, quantum antiferromagnets on arbitrary bi-partite or geometrically frustrated lattices, as well as Kondo lattice models. In these systems electrons of spin up and spin down are not allowed to simultaneously occupy certain lattice points. Remarkably, the partition function of such systems is naturally described by a new mathematical object --- the fermionant of a large matrix --- which is a simple generalization of the determinant. Interestingly, the fermionant also represents a particular $n$-point function of a system of free fermions. In particular, fermionants naturally appear when one sums over all Feynman diagrams of a given order in the perturbative expansion of the repulsive Hubbard model. The fermionant is mathematically related to the immanents of a matrix, for whose computation --- in contrast to the determinant --- no polynomial time algorithm is known. If a polynomial time algorithm for computing the fermionant would  exist, several physically relevant fermion sign problems could be solved. While this may not be the case, the relation between the fermionant and the partition function of these strongly correlated electron systems sheds new light on the computational complexity of the corresponding fermion sign problems.


We would like to thank H.-U.~Baranger, B.\ Charron, W.\ Detmold, P.\ deForcrand, M.\ Hastings, A.\ Li,  S.\ Mertens, L.~Mitas and B.\ M\"uller, K.\ Orginos, C.\ Riedtmann, and M.\ Troyer for interesting discussions. This work is supported by the DOE under contract DE-FG02-05ER41368 as well as by the Schweizerischer Nationalfonds. The Albert Einstein Center for Fundamental Physics at Bern University is supported by the cooperation project C-13 of the CRUS/SUK.

\bibliography{ref}

\end{document}